\documentclass[runningheads]{svmult}

\usepackage{makeidx}   
\usepackage{graphicx}  
\usepackage{subeqnar}  
\usepackage{multicol}  
\usepackage{physprbb}  
\makeindex             

\begin{document}
\title*{Intracluster Planetary Nebulae as Probes of 
Intracluster Starlight}
\toctitle{Intracluster Planetary Nebulae}
%
%
\titlerunning{Intracluster PN}
%
\author{John Feldmeier}
\authorrunning{John Feldmeier}
%
%
\institute{Case Western Reserve University, Cleveland OH, 44106, USA}

\maketitle              

\begin{abstract}

I review the progress in research on Intracluster Planetary Nebulae (IPN).
Hundreds of IPN candidates have now been found in the Virgo and Fornax
galaxy clusters, and searches of two nearby galaxy groups have made.  
From the results thus far, approximately 10--20\% of all 
stars in Virgo and Fornax are in an intracluster component, but
there are few such stars in galaxy groups.  From the spatial 
distribution of IPN, it appears that the intracluster stars are clustered,
in agreement with tidal-stripping scenarios.  In Virgo, the IPN have
a large line-of-sight depth, which implies that the bulk of intracluster
stars in this cluster derive from late-type galaxies and dwarfs.  
I also discuss other important developments in IPN research such 
as the detection of intracluster H~II regions, a possible 
detection of IPN in the Coma Cluster, and future observational and
theoretical developments.

\end{abstract}

\section{Introduction}

Why do planetary nebulae (PN), which are a late phase of stellar evolution, 
make excellent tracers of intracluster stars, stars between the
galaxies inside of a galaxy cluster?  An example of the utility
of PN under similar circumstances can be seen in Figure~1.  On the 
left is a standard broad-band image of the famous interacting 
galaxy pair M~51.  Although there are numerous signs of the 
interaction at high surface brightnesses, there are many 
lower surface-brightness features \cite{burk1978}  
that are barely visible in this representation. 

However, the distribution of PN found in M51 \cite{pnlf11} 
displayed on the right give additional information on 
this interacting system.  There is a clear tidal tail structure 
to the west of the secondary galaxy, and there are also 
signs of an extension of the spiral arm to the south.  The lack of 
PN in the central regions of the galaxies are due to crowding effects.  
What this illustration reminds us is that {\it planetary nebulae trace 
stellar luminosity, not stellar surface brightness}.  Because PN
are an end-phase of stellar evolution for most stars between
1--8 solar masses, the distribution of planetary nebulae closely
follows the distribution of stars in galaxies \cite{pnlf2}.  
If there is enough luminosity in a stellar population for sufficient 
numbers of PN to be present, then we can detect and study that population, 
regardless of its surface brightness.  Intracluster stars, which are 
believed to be quite luminous (anywhere between 10\% 
and 70\% of the cluster's total stellar luminosity; 
\cite{rich1983,miller1983}), but have a very low surface 
brightness (less than 1\% of the night sky in the optical bands; 
\cite{gon2000,icl1}) are therefore an ideal target for PN searches.

PN have additional features that make them useful probes of 
intracluster starlight.  Because PN are emission-line objects, 
any bright PN that can be detected photometrically 
can also be observed spectroscopically to high precision ($\sigma \approx
12$~km s$^{-1}$).  In the case of M~51, spectroscopic follow-up of 
the PN \cite{durr2003} found that the western tidal tail consists of 
two discrete structures that overlap in projection, one from each
galaxy.  With the high velocity precision of the PN observations, 
the separation of the two kinematic structures was trivial.  Although I will 
not focus on the dynamical aspects of IPN/PN in this
review  (see reviews by Arnaboldi, Douglas, Gerhard, \& Peng this conference), 
it is important to state that IPN are the easiest (and perhaps, 
the only) way to obtain dynamical information about the intracluster light.

Finally, through the [O~III] $\lambda$ 5007 Planetary Nebulae 
Luminosity Function (PNLF), extragalactic PN make excellent distance 
indicators (see Ciardullo, this conference).  The distinctive PNLF 
can be used in the intracluster environment to gain information on 
the line-of-sight distribution of the intracluster stars.    

Why study intracluster stars in the first place?  Once a curiosity
proposed by Zwicky (\cite{zwicky1951}), 
intracluster light is potentially of
great interest to studies of galaxy and galaxy cluster evolution.
The dynamical evolution of cluster galaxies is complex, and involves the 
poorly understood processes of cluster accretion and tidal stripping 
\cite{dressler1984}.  The intracluster light provides a unique way 
to study these mechanisms.  Modern numerical simulations show that
the intracluster light 
\cite{dub1998,murante2004,sommer2004,willman2004,mihos2004a} has
a complex structure, and can be used to gain information on 
the dynamical evolution of galaxies and galaxy clusters.        

\begin{figure}[t]
\begin{center}
\end{center}
\caption[]{Two images of the interacting galaxy pair M51.  On the left is
a broad-band image taken with the KPNO 4-m telescope showing the classical
high-surface brightness features of this interacting system.  
On the right shows the location of the 64 planetary nebula 
candidates found by \cite{pnlf11}.  Note the clear tidal tail 
structures that are extremely difficult to detect in the broad-band
image, but can be seen easily in the PN distribution.}
\label{fig:m51}
\end{figure}

\section{History of Intracluster Planetary Nebulae Research}

The history of IPN research begins over a decade ago with the 
first PN survey of the 
Virgo cluster \cite{pnlf5} (hereafter JCF).  In this
survey of elliptical galaxies, JCF found 11 PN 
that were much brighter than the expected [O~III] $\lambda$ 5007  
PNLF cut-off magnitude.  JCF attempted
to explain these ``overluminous'' PN with a number of hypotheses, 
but none was entirely satisfactory.

The next step involved spectroscopic follow-up of objects from the JCF
survey.  During a radial velocity survey of PN in the Virgo 
elliptical galaxy M~86, \cite{arna1996} found that 16 
of the 19 detected PN velocities were consistent with the galaxy's mean
velocity (v$_{radial}$ = -227 km s$^{-1}$).  The other three planetaries
had mean radial velocities of $\sim 1600$ km s$^{-1}$, more consistent with
the Virgo cluster's mean velocity.  \cite{arna1996} argued convincingly 
that these objects were intracluster planetary 
nebulae, and it is here that the term first enters the literature.  
The ``overluminous'' PN candidates were thus naturally explained as
a population of intracluster stars in front of the target galaxies.
Almost simultaneously, the first search for IPN candidates in
the Fornax cluster was published \cite{t1997}, and more 
detections of IPN candidates in Virgo quickly followed 
\cite{mendez1997,m87ipn,ipn1}.

However, a surprise was in the works.  Spectroscopic follow-up
of the IPN candidates revealed that some were not IPN, 
but instead background emission-line objects with extremely 
high equivalent width \cite{freeman2000,kud2000}.  This was unexpected,
because previous deep emission-line surveys had found very few
such objects \cite{pritchet1994}, though many have now been
detected at fainter magnitudes \cite{rhoads2000}.  The most likely source 
of the contamination was found to be Lyman-$\alpha$ galaxies at
redshifts 3.12--3.14, where the Lyman-$\alpha$ $\lambda$ 1215 line
has been redshifted into the [O~III] $\lambda$ 5007 filters used
in IPN searches.  However other types of contaminating objects
may also exist \cite{stern2000,norman2002}.  

Although these contaminants caused some consternation at first, 
a number of lines of evidence quickly showed
that the majority of IPN candidates are in fact, actual IPN.
Observations of blank control fields with identical search procedures
as the IPN surveys \cite{blank2002,castro2003} have 
found that the contamination fraction is significant, but was 
less than the observed IPN surface density.  The surface densities 
found correspond to a contamination rate of $\approx$ 20\% in the 
Virgo cluster and $\approx$ 50\% in Fornax (Fornax is more distant
than Virgo, so its PNLF is fainter, and therefore further down
the contaminating sources luminosity function).  There are still 
significant uncertainties in the background density due to 
large-scale structure, and to the small numbers of contaminating 
objects found thus far.  However,
deeper and broader control fields are forthcoming.  
Spectroscopic follow-up of IPN candidates 
\cite{freeman2000,blank2002,arna2003}, and
Arnaboldi et~al., this conference, clearly show large numbers of
IPN candidates have the expected [O~III] $\lambda$ 5007 and 4959
emission lines, with a contamination rate similar to the blank
field surveys.  Finally, there is independent evidence of 
individual intracluster stars in Virgo from observations of 
individual Intracluster Red Giant stars (IRGs; 
\cite{ftv1998,durr2002}) from the {\sl Hubble Space Telescope}.

It is worth reiterating that {\it all} deep [O~III] $\lambda$ 5007
emission-line searches will have such contamination at fainter
magnitudes.  In fact, such objects have already been seen 
in conventional PN surveys (\cite{mendez2001,durr2003} and
Romanowsky, private communication).  Extragalactic PN researchers
should keep this into account.  The brightest of such objects
have been detected with a $m_{5007}$ magnitude of 25.5 
\cite{freeman2000}, though the luminosity function of the contaminants
is still poorly known.

Currently, with the widespread use of mosaic CCD detectors, and
automated detection methods derived from DAOPHOT and SExtractor,
over a hundred IPN candidates can be found in a single telescope
run \cite{okamura2002,ipn2,arna2003,ipn3}.  IPN candidates are easily 
identified as stellar sources that appear in a deep [O~III] 
$\lambda$ 5007 image, but completely disappear in an image through a
filter that does not contain the [O~III] line.  Currently, over
400 IPN candidates have been detected in the Virgo cluster, and
over 100 IPN candidates have been found in the Fornax cluster.  
Figure~2 summarizes the status of the different surveys.  
Despite all of the effort, to date, only a few percent of 
the total angular area of Virgo and Fornax have been 
surveyed.  Literally thousands of IPN wait to be discovered by 
4-meter class telescopes.  

\begin{figure}[t]
\begin{center}
\end{center}
\caption[]{Images of the Virgo cluster (left), and the Fornax
cluster (right), with the various published 
IPN survey fields marked, as well as the two {\sl HST} IRG fields.  
Soon to come are additional fields from Aguerri et al. 
(2004; Virgo), Ciardullo et al. (2004; Fornax), Feldmeier 
et al. (2005; Virgo) and Ciardullo et al. (2005; IRG Virgo)}
\label{fig:map}
\end{figure}

\section{The Spatial Distribution of the Intracluster Light}

If we want to understand the mechanisms that produce intracluster 
stars, it is useful to compare the intracluster 
stars' spatial distribution with that
of the better-studied components of galaxy clusters: 
the cluster galaxies, the hot intracluster gas, and 
the invisible dark matter.  Theoretical work predicts very different
spatial distributions for the intracluster light depending on its
production time.  For example, if most intracluster stars are
removed early in a cluster's lifetime \cite{mer1984}, the distribution
of this diffuse component will be smooth and follow the cluster potential.  
However, if a significant portion of the stars are removed at 
late epochs via galaxy encounters and tidal stripping 
\cite{rich1983,harass}, then the intracluster light should be clumpy 
and have a non-relaxed appearance.  In particular,
the ``harassment'' models of \cite{harass,mlk1998} first predicted
that many intracluster stars will exist in long ($\sim 2$~Mpc) tidal
tails, which may maintain their structure for Gyrs.

With regards to angular clustering, there is good evidence that
the IPN follow a non-random distribution.  \cite{okamura2002,ipn3}
show evidence for clumps of IPN within their survey fields, and 
Aguerri et al. (2004) found a strong clustering signal using a 
standard two-point correlation function analysis 
(reported in \cite{nap2003}).  The scale of the angular clustering 
ranges from one to ten arcminutes, (corresponding 5 to 50 kpc assuming
a mean Virgo distance of 15 Mpc), depending on the field studied.  
However, no tail-like features have yet been seen in the 
IPN distribution, although tidal arc features have been seen in 
broad-band images of other galaxy clusters 
\cite{tren1998,gregg1998,cr2000}.

\begin{figure}[t]
\begin{center}
\includegraphics[width=1.0\textwidth]{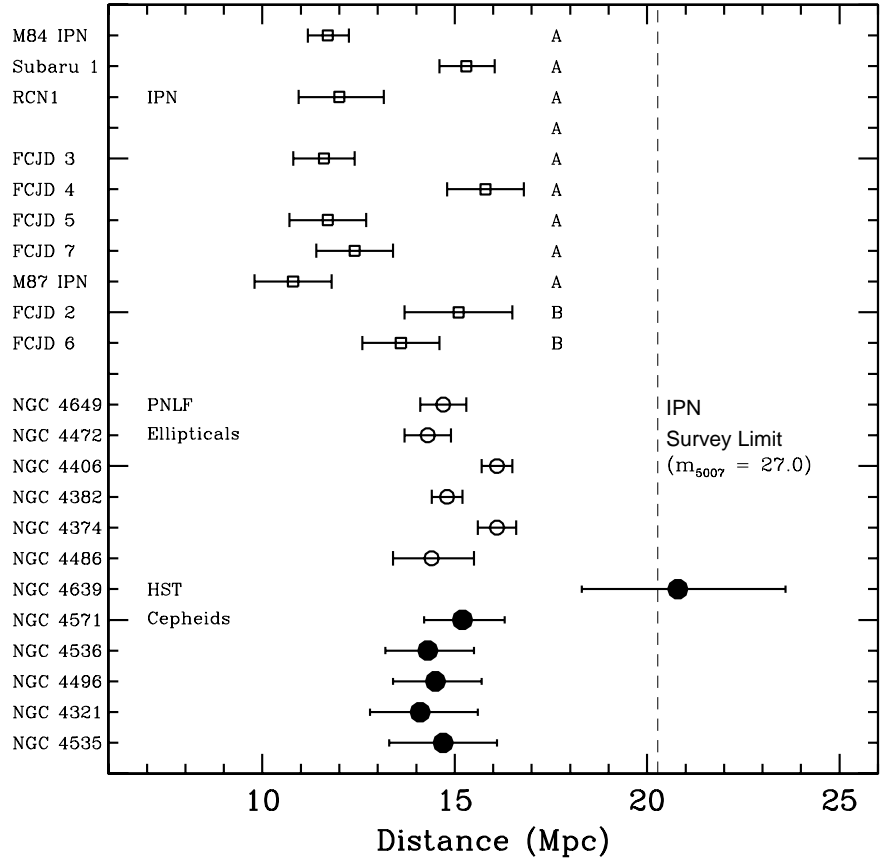}
\end{center}
\caption[]{A comparison of the upper limit distances obtained from the  
intracluster planetary nebulae to direct distances to Virgo
Cluster galaxies.  At the top are the upper limit distances (denoted 
by the open
squares) from IPN observations
by Okamura et al. (2002) and Arnaboldi et al. (2002).
Below that are the distances derived from as the 
overluminous IPN found in front of M87 (Ciardullo et al. 
1998).  These upper limit distances are compared to the PNLF distances of 
Virgo ellipticals (denoted by the open circles; 
Jacoby, Ciardullo, \& Ford 1990; Ciardullo et al. 1998), 
and Cepheid distances to spiral galaxies 
(denoted by the filled circles; Pierce et al. 1994; Saha et al. 1997; 
Freedman et al. 2001).  The subcluster of Virgo that each 
intracluster field resides in is noted.  Note the great depth of
IPN, compared to that of the elliptical galaxies.}
\label{fig:depth}
\end{figure}

With the use of the [O~III] $\lambda$ 5007 PNLF, we can also obtain
line-of-sight information on the intracluster stars, a property
unique to IPN.  Although a full analysis of the observed luminosity
functions has not yet been attempted, we can already learn something 
useful from the brightest IPN candidates in each field.  Given the empirical 
PNLF, there is a maximum absolute magnitude $M^*$, 
that a PN may have in the light of [O~III].  If we assume that the 
brightest observed IPN candidate in each field has this absolute magnitude,
we can derive an upper limit on its distance, and therefore estimate
the distance to the front edge of the Virgo intracluster
population.  This is plotted in Figure~\ref{fig:depth}.
As can be clearly seen, there is an offset between the IPN of
subclusters A and B : this is interpreted as due to the differing 
distances of these two subclusters.  However, the most revealing feature
of the distribution of the IPN is in subclump A.  The upper limit
distance for most of the fields is well in front of the cluster
core.  This is partially due to a selection effect: it is easier
to detect an IPN on the near side of Virgo than the far side.
Nevertheless, the depth implied from the measurements is remarkable.  
If we take the data at face value, then the IPN distribution 
has a line-of-sight radius of over 4~Mpc.  If we compare this radius 
to the classical radius of Virgo on the sky (six degrees, or 1.6 Mpc), 
we find that the Virgo cluster is more than 2.6 times as deep as 
it is wide.  Virgo is nowhere near a spherical cluster: it contains 
considerable substructure, and is elongated significantly along 
our line of sight.

The great depth derived for the IPN distribution gives an important clue
to the parent galaxies of the intracluster stars.  Numerical simulations 
\cite{mlk1998,dub2000} show that the majority of
intracluster stars are ejected into orbits similar to that of their parent
galaxies.  In Virgo, the cluster ellipticals are clustered within
a radius of $\approx 2$ Mpc from the cluster core \cite{pnlf5}.
In contrast, the hydrogen deficient spirals found in the Virgo
cluster have a radius of 4 Mpc or larger \cite{sanchis2002}, and
the dwarf ellipticals have a depth of 6 Mpc \cite{jerjen2003}.
Therefore, it seems clear that a significant portion of 
Virgo's intracluster stars originate from late-type galaxies 
whose highly radial orbits take them in and out of the cluster core.

\section{Converting IPN densities to Luminosity Densities}

In principle, determining the amount of intracluster luminosity from 
the observed numbers of IPN is straightforward.  Theories of simple 
stellar populations \cite{renzini} have shown that the 
bolometric luminosity-specific stellar evolutionary flux 
of non-star-forming stellar populations should 
be $\sim 2 \times 10^{-11}$~stars-yr$^{-1}$-$L_{\odot}^{-1}$, (nearly) 
independent of population age or initial mass function.  If the 
lifetime of the planetary nebula stage is $\sim 25,000$~yr,
and if the empirical PNLF is valid to $\sim 8$~mag below 
the PNLF cutoff, then every stellar system should have $\alpha \sim 50 \times 
10^{-8}$~PN-$L_{\odot}^{-1}$.  According to the empirical PNLF,
approximately one out of ten of these PNe will be within
2.5~mag of $M^*$.  Thus, under the above assumptions,
most stellar populations should have   
$\alpha_{2.5} \sim 50 \times 10^{-9}$~PN-$L_{\odot}^{-1}$.  
The observed number of IPN, coupled with the PNLF, 
can therefore be used to deduce the total luminosity
of the underlying stellar population.

In practice, there are a number of systematic effects that must be accounted
for before we can transform the numbers of IPN to a stellar luminosity
\cite{ipn3}, which we briefly summarize here.

First, although stellar evolution theory originally predicted a constant 
$\alpha_{2.5}$ value for all non star-forming populations, 
observations present a more complicated picture.  
\cite{rbciau} found that in a sample 
of 23 elliptical galaxies, lenticular galaxies, and spiral bulges, 
the observed value of $\alpha_{2.5}$ never exceeded 
$\alpha_{2.5} = 50 \times 10^{-9}$~PN-$L_{\odot}^{-1}$ but was 
often significantly less, with higher luminosity galaxies 
having systematically smaller values of $\alpha_{2.5}$.  Since the 
amount of intracluster starlight derived is inversely proportional
to the $\alpha_{2.5}$ parameter, a large error in the amount of
intracluster light can result if this is not accounted for.

By comparing the numbers of IPN in a field surrounding a 
{\sl HST} WFPC2~field, with RGB and AGB star counts, 
\cite{durr2002} found a value of $\alpha_{2.5} = 23^{+10}_{-12} 
\times 10^{-9}$~PN-$L_{\odot}^{-1}$ for Virgo's intracluster population.  
This observational measurement is bolstered by the new theoretical
work of Buzzoni \& Arnaboldi (this conference), which show that
the observational value for $\alpha_{2.5}$ is well within the
range for models of moderate luminosity galaxies.

Second, as we have discussed previously, IPN surveys are not pristine:
approximately 20\% of Virgo IPN candidates, and 50\% of Fornax IPN
candidates, are likely to be unrelated background sources.  Although
these objects are subtracted out statistically from all modern IPN
surveys, the importance of field-to-field variations in the background
due to large-scale structure is still unknown.    

Finally, the IPN candidates of Virgo have a significant line-of-sight 
depth.  Since the conversion between number of PN and luminosity
depends on the shape of the luminosity function, this depth can
change the amount of intracluster light found from the data.
Models \cite{ipn1} indicate that the difference between a 
single-distance model (the most conservative) and assuming that
the IPN are uniformly distributed in a sphere of radius 3 Mpc (the least
conservative), changes the derived intracluster star luminosity 
by up to a factor of three.  Thus far, all IPN researchers have 
adopted a single distance model in order to be conservative, but 
this effect is the least studied at this point.

After applying all of these corrections, the intracluster stellar
fractions for Virgo
vary between 10 and 20\% \cite{rcn1,okamura2002,arna2003,ipn3},
with the errors being dominated by the systematic effects.  The
IRG measurements \cite{ftv1998,durr2002} find somewhat less intracluster
light (10--15\%), but the various results agree within the errors.
However, it is important to note that
each of these studies makes different assumptions concerning
the calculation of intracluster luminosity, the amount of
light bound to galaxies, and the contamination of background sources.
For Fornax, there are only two measurements of the intracluster
star fraction thus far from IPN.  
\cite{t1997} report a fraction of up to 40\%, but
this is before the detection of contaminating sources, and is likely
to be an overestimate.  Ciardullo (2004, in prep), reports an
approximate fraction of 20\%, with similar errors as the Virgo
results.  Regardless of the uncertainties, it is clear that a significant
fraction of all stars in Virgo and Fornax are in an intracluster 
component.  

How does the IPN luminosity density compare with well-known cluster
properties such as radius, or projected galaxy density?  \cite{ipn3}
has compared these properties in Virgo, and has found little or no 
correlation.  This may be due to Virgo's status as a dynamically young
cluster, or due to a selection effect (IPN researchers must observe
where the galaxies are not, to avoid confusion with normal extragalactic
PN).  More data will be needed to confirm this result.

\section{Intracluster H~II Regions}

A recent discovery stemming from IPN searches is the
detection of H~II regions in the far halos of galaxies 
or in intracluster space.  In the course of spectroscopic 
follow-up of a Virgo IPN field, \cite{gerhard2002} 
found an emission-line source that
has the properties of a $\sim$ 400 solar mass, 3 Myr compact
H~II region, over 17 kpc away from the nearest Virgo galaxy
(see also \cite{lee2000} for a similar, earlier, under-appreciated 
example).  There are now a number of other examples of 
intergalactic star formation in Virgo and in other 
galaxy clusters and groups 
\cite{xu1999,sakai2002,ryan-weber2003,cortese2004,durr2004}.

These discoveries have two important implications.  First, star formation
can occur at large distances from galaxies, and environments quite
different than that normally studied.  This has implications for
a number of fields of astrophysics, including the origin of 
metallicity in the intergalactic and intracluster medium, and 
the possible in situ origin of B stars in the Galactic halo.  At this
time, it does appear that intracluster star formation is only a small
fraction of the total intracluster star production, but the exact
amount is uncertain.  Second, these objects are another form 
of contamination to pure IPN 
surveys.  Discouragingly, the only way to separate these 
objects from IPN is through deep 
spectroscopy.  Luminous PN have [O~III] / H$\alpha$ ratios of two
or greater \cite{pnlf12}, where luminous H~II regions do not.  
From the surveys thus far, these objects appear to be relatively
rare component of IPN surveys ($\sim 3\%$), but more study is 
needed in this area.

\section{Intra-group Starlight}

Although the presence of intracluster starlight in clusters such as
Virgo and Fornax is now well established, the amount of 'intra-group'
starlight is still uncertain.  Theoretical studies predict that 
if most intracluster stars are removed by galaxy collisions 
\cite{rich1983,harass}, the fraction of intra-group stars, 
to first order, should be a smooth function of galaxy number 
density (L$_{{ICL}}$ $\sim$ N$_{{Gal}}^{2}$).  To test 
this hypothesis, two different groups have surveyed the nearby
M~81 and Leo groups of galaxies \cite{castro2003,durr2004}, using similar
detection methods as the cluster searches.  In both cases, no
genuine intra-group PN were found.  These non-detections
strongly implies there is substantially less (4--15 times) less 
intergalactic stars in groups than there are in clusters.  

When compared with other measurements of intracluster star fractions    
through modern deep imaging \cite{bern1995,gon2000,icl1,icl2}, 
or through the detection of intracluster
supernovae \cite{gal-yam2003}, an interesting pattern emerges, plotted
in Figure~\ref{fig:fraction}.  For
galaxy clusters, the data is consistent with an approximate fraction of
20\%, albeit with large error bars or intrinsic scatter. 
 When we move to the group environment, the fraction abruptly drops, 
with no sign of any smooth decline.  This implies that there is 
something special about the cluster environment that promotes 
intracluster star production.  More data will be needed to confirm
and strengthen this result, especially IPN searches of 
additional galaxy groups.  

\begin{figure}[t]
\begin{center}
\end{center}
\caption[]{A comparison of detected intracluster star fractions from modern
measurements, as a function of the velocity dispersion.  Note the abrupt
change from the cluster environment to the group environment.}
\label{fig:fraction}
\end{figure}

\section{IPN in the Coma Cluster?}

At this conference, Gerhard et al. announced the possible detection of 
at least one IPN candidate in the Coma Cluster.  If this 
candidate is confirmed, it will be the most distant individual 
PN ever detected, at a distance of over 100 Mpc.  This object 
was found using a technique of filling the entire focal plane of 
the 8-m Subaru telescope with a slit mask through the appropriate
[O~III] $\lambda$ 5007 filter, and looking for narrow-line 
emission sources.  Deep broad-band imaging has implied that the Coma Cluster 
may have a very high intracluster star fraction, up to 50\% 
\cite{bern1995}, so it is plausible that a few IPN could be detected, 
despite the small area surveyed.  The most exciting aspect 
of this observation is that it opens up a way to observe PN at 
much greater distances than previously thought possible, and 
makes it possible to place IPN density limits in more distant 
galaxy clusters.    

\section{The Future}

Studies of IPN are not even a decade old, and much more work 
still needs to be done.  The most crucial observations needed 
in the near term are spectroscopic follow-up of a large number 
of IPN candidates.  This would allow us to better determine 
the contamination rate  from background galaxies, 
avoid any other intracluster H~II regions, and most importantly,
gain information on the kinematics of intracluster stars.
More IPN imaging is needed in Virgo and Fornax at differing
radii and densities, in order to confirm the lack of correlations
found thus far.  IPN surveys of additional galaxy groups are
needed to determine the lack of intra-group stars in these systems,
and to better characterize the steepness of the intracluster 
production ``cliff.''

However, the future of IPN research will involve close comparisons
of IPN properties to other observations of extragalactic PN and  
galaxy clusters and close comparison to the modern numerical 
simulations now being undertaken.  Some examples are given below, 
but it is almost certain that many more will be added in the next few years.

Villaver (this conference), presented the first numerical hydrodynamical
simulations of IPN within a hot intracluster gaseous medium.  The
results imply that IPN do survive in the intracluster medium, but the 
nebula may be spatially distorted. 

\cite{mihos2004b} has just completed the first phase of a 
deep CCD imaging survey of the Virgo cluster core using the 
techniques of ultra-deep surface photometry.  These new observations
reach extremely faint surface brightnesses ($\mu_{V} \approx 29.6$)
over large angular areas ($\approx 2$ square degrees).
Comparing these deep imaging maps with the IPN observations should
be revealing, and allow us to better determine the reality
of tidal features in both data sets, and to give clear targets
for follow-up IPN observations.  

{\sl HST}+ACS broad-band observations are planned of an intracluster
field in Virgo by Summer 2005, with the goal of obtaining the first 
color-magnitude
diagram of the intracluster stars.  By placing tighter constraints 
on the age and luminosity of the intracluster population, we 
should be able to better constrain the $\alpha_{2.5}$ parameter, 
and improve the accuracy of the total amount of intracluster starlight.
 
Finally, with the advent of advanced numerical simulations of galaxy
clusters, we will be able to compare the IPN data to models of
intracluster star production.  Due to the large angular size 
of nearby clusters, it will be many
years before the majority of the clusters will be surveyed for IPN.  
Numerical simulations will hopefully allow us to survey more efficiently, 
and to obtain better quality information on the properties of 
intracluster starlight.

\null\par
I would like to thank the conference organizers for allowing me to give
this review, and for running an excellent conference.  I would also 
like to thank my collaborators for their years of effort on the science 
presented here.  This work was supported in part by NSF grant 
AST 03-02030 and NASA grant NAG 5-9377.

\null\par

%


\begin{thebibliography}{8.}
\addcontentsline{toc}{section}{References}

\bibitem{arna1996} Arnaboldi, M., et al.
1996, Ap.~J., 472, 145

\bibitem{rcn1} Arnaboldi, M., et al.\ 
2002, A.~J., 123, 760 

\bibitem{arna2003} Arnaboldi, M., et al.\ 
2003, A.~J., 125, 514 

\bibitem{bern1995} Bernstein, G. M., et al.\ 1995, A.~J., 110, 1507

\bibitem{burk1978} Burkhead, M.~S.\ 1978, Ap.~J.~Suppl., 38, 147

\bibitem{cr2000} Calc\'aneo-Rold\'an, C., et al.\ 2000, M.N.R.A.S., 314, 324

\bibitem{castro2003} 
Castro-Rodr{\'{\i}}guez, N., et al.\ 2003, Astr.~Ap.~405, 803 

\bibitem{rbciau}
Ciardullo, R. 1995, in IAU Highlights of Astronomy, 10, 
ed.~I. Appenzeller (Dordrecht: Kluwer), p.~507

\bibitem{pnlf2}
Ciardullo, R., et al.\ 1989, Ap.~J., 339, 53

\bibitem{m87ipn} Ciardullo, R., et al.\ 1998, Ap.~J., 492, 62

\bibitem{blank2002} Ciardullo, R., et al.\ 2002a, Ap.~J., 566, 784

\bibitem{pnlf12}
Ciardullo, R., et al.\ 2002b, Ap.~J., 577, 31

\bibitem{cortese2004} Cortese, L., et al. \ 2004, A.\&~A., 416, 119

\bibitem{dressler1984}Dressler, A. 1984, A.R.A.\&~A., 22, 185

\bibitem{dub1998}Dubinski, J. 1998, Ap.~J., 502, 141

\bibitem{dub2000} Dubinski, J., Murali, C.,\& Ouyed, R. 
2000, unpublished preprint

\bibitem{durr2002} Durrell, P.~R., et al. \ 2002, Ap.~J., 570, 119

\bibitem{durr2003} Durrell, P.~R., et al. \ 2003, Ap.~J., 582, 170 

\bibitem{durr2004} Durrell, P.~R., et al. \ 2004, IAU Symposium, 217, 90 

\bibitem{pnlf11} 
Feldmeier, J.~J., Ciardullo, R., \& Jacoby, G.~H.\ 1997, Ap.~J., 479, 231

\bibitem{ipn1} 
Feldmeier, J. J., Ciardullo, R., \& Jacoby, G. H. 1998, Ap.~J., 503, 109 

\bibitem{icl1} Feldmeier, J. J., et al. 2002, Ap.~J., 575, 779

\bibitem{ipn2} 
Feldmeier, J. J., et al. \ 2003a, Ap.~J.Suppl., 145, 65

\bibitem{icl2} Feldmeier, J. J., et al. 2004a, Ap.~J., in press

\bibitem{ipn3} 
Feldmeier, J. J., et al. \ 2004b, Ap.~J., in press

\bibitem{ftv1998} 
Ferguson, H. C., Tanvir, N. R., \& von Hippel, T. 1998, Nature, 391, 461 

\bibitem{freeman2000} Freeman, K.~C.~et al.\ 
2000, ASP Conf.~Ser.~197: Dynamics of Galaxies: from the Early Universe to 
the Present, ed. F. Combes, G. A. Mamon, \& V. Charmandaris 
(San Francisco: ASP), 389.

\bibitem{gal-yam2003} Gal-Yam, A., et al. \ 2003, A~J., 125, 1087 

\bibitem{gerhard2002} Gerhard, O., et al. \ 2002, Ap.~J.~L., 580, L121 

\bibitem{gon2000} Gonzalez, A. H., et al. \ 2000, Ap.~J., 536, 561

\bibitem{gregg1998} Gregg, M. D., \& West, M. J. 1998, Nature, 396, 549

\bibitem{pnlf5} Jacoby, 
G.~H., Ciardullo, R., \& Ford, H.~C.\ 1990, Ap.~J., 356, 332 

\bibitem{jerjen2003} Jerjen, H., Binggeli, B. \& Barazza, F.D. 2004, A.J., 
127, 771

\bibitem{kud2000} 
Kudritzki, R.-P., et al. \ 2000, Ap.~J., 536, 19

\bibitem{lee2000} Lee, H., 
Richer, M.~G., \& McCall, M.~L.\ 2000, Ap.~J.L., 530, L17 

\bibitem{mendez1997} M\'endez, R. H., et al. \ 1997, Ap.~J.~L., 491, 23

\bibitem{mendez2001} M{\' e}ndez, R.~H., \ 2001, Ap.~J., 563, 135

\bibitem{mer1984} Merritt, D. 1984, Ap.~J., 276, 26

\bibitem{mihos2004a} Mihos, J.C., et al. 2004a, in preparation

\bibitem{mihos2004b} Mihos, J.C., et al. 2004b, in preparation

\bibitem{miller1983} Miller, G.E. 1983, Ap.~J., 268, 495

\bibitem{harass} 
Moore, B., et al., 1996, Nature, 379, 613

\bibitem{mlk1998} Moore, B., Lake, G. \& 
Katz, N. 1998, Ap.~J., 495, 139

\bibitem{murante2004} Murante, G., et al.\ 
2004, Ap.~J.~L., 607, L83 

\bibitem{nap2003} Napolitano, N.~R., et al.\ 2003, Ap.~J., 594, 172 

\bibitem{norman2002} Norman, C., et al.\ 2002, Ap.~J., 571, 218

\bibitem{okamura2002} Okamura, S., et al. 2002, P.A.S.J., 54, 883

\bibitem{pritchet1994} Pritchet, C.~J.\ 1994, P.A.S.P., 106, 1052

\bibitem{renzini} A. Renzini, A. Buzzoni: In: \emph{Spectral Evolution of 
Galaxies}, ed.~by C. Chiosi, A.~Renzini (Reidel, Dordrecht), p.~195 (1986)

\bibitem{rhoads2000} Rhoads, J.~E., et al. \ 2000, Ap.~J.~L., 545, L85

\bibitem{rich1983} 
Richstone D. O., \& Malumuth, E. M. 1983, Ap.~J., 268, 30

\bibitem{ryan-weber2003}
Ryan-Weber, E. V.,  et~al. 2004, A.~J., 127, 1431

\bibitem{sakai2002} Sakai, S., Kennicutt, R.~C., van der 
Hulst, J.~M., \& Moss, C.\ 2002, Ap.~J., 578, 842

\bibitem{sanchis2002} Sanchis, T., et al. \ 2002, Ap~J., 580, 164 

\bibitem{sommer2004}  Sommer-Larsen, J., Romeo, A.D., Portinari, L. 2004,
M.N.R.A.S., submitted, available as astro-ph/0403282

\bibitem{stern2000} Stern, 
D., Bunker, A., Spinrad, H., \& Dey, A.\ 2000, Ap.~J., 537, 73

\bibitem{t1997} Theuns, T., \& 
Warren, S. J. 1997, M.N.R.A.S., 284, L11

\bibitem{tren1998} Trentham, N. \& Mobasher, B. 1998, M.N.R.A.S., 293, 53

\bibitem{willman2004} Willman, B., et al. 2004, M.N.R.A.S., submitted - 
available as astro-ph/0405094

\bibitem{xu1999} Xu, C., 
Sulentic, J.~W., \& Tuffs, R.\ 1999, Ap.~J., 512, 178 

\bibitem{yun1994} Yun, M.S., Ho, P.T.P., \& Lo, K.Y. 1994, Nature, 
372, 530

\bibitem{zwicky1951} Zwicky, F. 1951, P.A.S.P., 63, 61
\end{thebibliography}
\end{document}